\documentclass[superscriptaddress,twocolumn,showpacs,preprintnumbers,amsmath,amssymb,floatfix,aps]{revtex4}

\usepackage[usenames]{color}
\usepackage{placeins}

\usepackage{graphicx}

\ifx\pdfoutput\undefined
  \usepackage{epsfig}
  \epsfclipon
\fi

\newcommand{\MeVc}{\text{MeV}} 
\newcommand{\GeVc}{\text{GeV}} 

\begin{document}

\title{A Search for $\Sigma^0_5$, $\text{N}^0_5$ and $\Theta^{++}$ Pentaquark States}

\author{Y.~Qiang}
\affiliation{Massachusetts Institute of Technology, Cambridge, MA 02139, USA}

\author{J.~Annand}
\affiliation{University of Glasgow, Glasgow, Scotland}

\author{J.~Arrington}
\affiliation{Physics Division, Argonne National Laboratory, Argonne, IL 60439, USA}

\author{Ya.I.~Azimov}
\affiliation{Nuclear Physics Institute, St. Petersburg, Russia 188300}

\author{W. Bertozzi}
\affiliation{Massachusetts Institute of Technology, Cambridge, MA 02139, USA}

\author{G.~Cates}
\affiliation{Department of Physics, University of Virginia, Charlottesville, VA 22904, USA}

\author{J.~P.~Chen}
\affiliation{Thomas Jefferson National Accelerator Facility, Newport News, VA 23606, USA}

\author{Seonho~Choi}
\affiliation{Temple University, Philadelphia, PA 19122, USA}

\author{E.~Chudakov}
\affiliation{Thomas Jefferson National Accelerator Facility, Newport News, VA 23606, USA}

\author{F.~Cusanno}
\affiliation{INFN/Sanita, Rome, Italy}

\author{C.W.~de~Jager}
\affiliation{Thomas Jefferson National Accelerator Facility, Newport News, VA 23606, USA}

\author{M.~Epstein}
\affiliation{California State University, Los Angeles, Los Angeles, CA 90032, USA}

\author{R.J.~Feuerbach}
\affiliation{Thomas Jefferson National Accelerator Facility, Newport News, VA 23606, USA}

\author{F.~Garibaldi}
\affiliation{INFN/Sanita, Rome, Italy}

\author{O.~Gayou}
\affiliation{Massachusetts Institute of Technology, Cambridge, MA 02139, USA}

\author{R.~Gilman}
\affiliation{Rutgers, The State University of New Jersey, Piscataway, NJ 08854  USA}
\affiliation{Thomas Jefferson National Accelerator Facility, Newport News, VA 23606, USA}

\author{J.~Gomez}
\affiliation{Thomas Jefferson National Accelerator Facility, Newport News, VA 23606, USA}

\author{D.J.~Hamilton}
\affiliation{University of Glasgow, Glasgow, Scotland}

\author{J.-O.~Hansen}
\affiliation{Thomas Jefferson National Accelerator Facility, Newport News, VA 23606, USA}

\author{D.W.~Higinbotham}
\affiliation{Thomas Jefferson National Accelerator Facility, Newport News, VA 23606, USA}

\author{T.~Holmstrom}
\affiliation{College of William and Mary, Williamsburg, VA 23187, USA}

\author{M.~Iodice}
\affiliation{INFN/Roma Tre, Rome, Italy}

\author{X.~Jiang}
\affiliation{Rutgers, The State University of New Jersey, Piscataway, NJ  08854  USA}

\author{M.~Jones}
\affiliation{Thomas Jefferson National Accelerator Facility, Newport News, VA 23606, USA}

\author{J.~LeRose}
\affiliation{Thomas Jefferson National Accelerator Facility, Newport News, VA 23606, USA}

\author{R.~Lindgren}
\affiliation{Department of Physics, University of Virginia, Charlottesville, VA 22904, USA}

\author{N.~Liyanage}
\affiliation{Department of Physics, University of Virginia, Charlottesville, VA 22904, USA}

\author{D.J.~Margaziotis}
\affiliation{California State University, Los Angeles, Los Angeles, CA 90032, USA}

\author{P.~Markowitz}
\affiliation{Florida International University, Miami, FL 33199, USA}

\author{V.~Mamyan}
\affiliation{Yerevan Physics Institute, Yerevan, Armenia}

\author{R.~Michaels}
\affiliation{Thomas Jefferson National Accelerator Facility, Newport News, VA 23606, USA}

\author{Z.-E.~Meziani}
\affiliation{Temple University, Philadelphia, PA 19122, USA}

\author{P.~Monaghan}
\affiliation{Massachusetts Institute of Technology, Cambridge, MA 02139, USA}

\author{C.~Mu\~{n}oz-Camacho}
\affiliation{SACLAY, Gif-Sur-Yvette, France}

\author{V.~Nelyubin}
\affiliation{Department of Physics, University of Virginia, Charlottesville, VA 22904, USA}

\author{K.~Paschke}
\affiliation{University of Massachusetts, Amherst, 01003 MA, USA}

\author{E.~Piasetzky}
\affiliation{Tel Aviv University, Tel Aviv  69978, Israel}

\author{I.~Rachek}
\affiliation{Budker Institute of Nuclear Physics, Novosibirsk, Russia}

\author{P.E.~Reimer}
\affiliation{Physics Division, Argonne National Laboratory, Argonne, IL 60439, USA}

\author{J.~Reinhold}
\affiliation{Florida International University, Miami, FL 33199, USA}

\author{B.~Reitz}
\affiliation{Thomas Jefferson National Accelerator Facility, Newport News, VA 23606, USA}

\author{R.~Roche}
\affiliation{Florida State University, Tallahassee, FL, 32306  USA}

\author{A.~Saha}
\affiliation{Thomas Jefferson National Accelerator Facility, Newport News, VA 23606, USA}

\author{A.J.~Sarty}
\affiliation{Saint Mary's University, Halifax, NS, B3H 3C3, Canada}

\author{E.~Schulte}
\affiliation{Physics Division, Argonne National Laboratory, Argonne, IL 60439, USA}

\author{A.~Shahinyan}
\affiliation{Yerevan Physics Institute, Yerevan, Armenia}

\author{R.~Sheyor}
\affiliation{Tel Aviv University, Tel Aviv  69978, Israel}

\author{J.~Singh}
\affiliation{Department of Physics, University of Virginia, Charlottesville, VA 22904, USA}

\author{I.I.~Strakovsky}
\affiliation{The George Washington University, Washington, DC 20052, USA}

\author{R.~Subedi}
\affiliation{Kent State University, Kent, OH 44242, USA}

\author{R.~Suleiman}
\affiliation{Massachusetts Institute of Technology, Cambridge, MA 02139, USA}

\author{V.~Sulkosky}
\affiliation{College of William and Mary, Williamsburg, VA 23187, USA}

\author{B.~Wojtsekhowski}
\affiliation{Thomas Jefferson National Accelerator Facility, Newport News, VA 23606, USA}

\author{X.~Zheng}
\affiliation{Physics Division, Argonne National Laboratory, Argonne, IL 60439, USA}

\collaboration{and the Jefferson Lab Hall A Collaboration}
\noaffiliation

\date{\today}

\begin{abstract}
A high-resolution ($\sigma_{\text{instr.}} = 1.5~\MeVc$) search for
narrow states ($\Gamma < 10~\MeVc$) with masses of $M_x\approx
1500-1850~\MeVc$ in $ep \rightarrow e^\prime K^+X$, $e^\prime K^-X$
and $e^\prime \pi^+X$ electroproduction at small angles and low $Q^2$
was performed.  These states would be candidate partner states of the
reported $\Theta^+(1540)$ pentaquark.  No statistically significant
signal was observed in any of the channels at 90\% C.L.  Upper limits
on forward production were determined to be between 0.8\% and 4.9\% of
the $\Lambda(1520)$ production cross section, depending on the channel
and the assumed mass and width of the state.
\end{abstract}

\maketitle

The discovery of exotic baryonic states with positive strangeness,
requiring a minimal configuration of four quarks and an antiquark,
would contribute greatly to the understanding of confinement in
Quantum Chromodynamics (QCD).  Although searches for such states have
been conducted for almost 40 years with both partial wave analyses of
hadroproduction ({\it e.g.}~\cite{PhysRevLett.17.102}) and
electroproduction ({\it e.g.}~\cite{Mori:1970ne}), these early results
have generally been interpreted as unconvincing~\cite{Trippe:1976aq}.
Recent claims of the observation of one such state, the
$\Theta^+(1540)$~\cite{Nakano:2003qx} have generated renewed
experimental and theoretical interest in this topic.  For recent
reviews of the experimental evidence, see
Refs.~\cite{Hicks:2005gp,Burkert:2005ft,Schumacher:2005wu}. If
confirmed, the $\Theta^+$ could be the lowest-mass member of an
antidecuplet of pentaquark states, predicted within the framework of
the Chiral Quark Soliton Model~\cite{Diakonov:1997mm}.  Alternatively,
such exotic baryons have been explained in terms of models based on
diquark configurations~\cite{Jaffe:2003sg}, or in terms of
isospin-violating strong decays, which lead to an isotensor multiplet
of $\Theta$-pentaquarks of different charge
states~\cite{Capstick:2003iq}.  If the $\Theta^+$ pentaquark exists
then other members of its symmetry group and/or other multiplets
containing exotic states~\cite{Arndt:2003ga,Azimov:2005hv} should be
observable as well, provided they are sufficiently narrow.  All of the
approaches mentioned predict partner states in the mass region
$M\approx 1500-2000~\MeVc$.

This paper reports on a high-resolution search at forward production
angles for the $\Sigma^0_5$ and $\text{N}^0_5$ non-exotic members of
the antidecuplet and for the exotic $\Theta^{++}$ as narrow resonances
in the missing mass spectra of the reactions $ep\rightarrow eK^+X$,
$ep\rightarrow e\pi^+X$ and $ep\rightarrow eK^-X$, respectively.  The
measurements covered a limited range of small scattering angles, which
did not allow a partial-wave analysis; however, both experimental
indications~\cite{Nakano:2003qx} and theoretical
expectations~\cite{Kwee:2005dz} are for the $\Theta^{+}$ cross section
to be forward peaked.  One of the interesting features of the reported
observations of the $\Theta^+$ is that the measured width was either
an upper limit from the experimental resolution or consistent with
having a negligible width~\cite{Hicks:2005gp}.  Based on this, the
present measurement was specifically designed to be able to observe
and determine the width of extremely narrow states.  At large
scattering angle, a recently reported search found no statistically
significant evidence for the $\Theta^{++}$ ~\cite{Kubarovsky:2006ns}.
In the present experiment, very good mass resolution was
achieved. Precise measurements of the known $\Lambda(1116)$,
$\Sigma(1193)$ and $\Lambda(1520)$ states were obtained for
calibration.

The experiment took place in Hall A at Thomas Jefferson National
Accelerator Facility (Jefferson Lab) using a $5~\GeVc$ electron beam
incident on a 15~cm liquid hydrogen target. Scattered electrons were
detected in one of the High-Resolution Spectrometers
(HRS)~\cite{Alcorn:2004sb} in coincidence with electroproduced hadrons
in the second HRS.  Each spectrometer was positioned at $6^\circ$
relative to the electron beam by using a septum~\cite{Brindza_septa}
magnet to achieve this small scattering angle. The spectrometers had
an effective acceptance of approximately 4~msr in solid angle and $\pm
4.5\%$ in momentum. To obtain the desired missing mass coverage, the
central momentum of the electron HRS was varied between 1.85 and
2.00~$\GeVc$, while the central momentum of the hadron HRS was changed
between 1.89 and 2.10~$\GeVc$.  In these configurations, the average
momentum transfer of the virtual photon was $\langle Q^2\rangle\approx
0.1~\GeVc^2$, and the average center-of-mass (CM) photon energy was
$\langle E^\text{CM}_\gamma\rangle = 1.1$~GeV.  For the kaon (pion)
kinematics, the center-of-mass scattering angle was $5.6^\circ \le
\theta^\text{CM}_{\gamma^*K} \le 11.4^\circ$ ($5.0^\circ \le
\theta^\text{CM}_{\gamma^*\pi} \le 10.4^\circ$), and the angular
acceptance was $\Delta\Omega^\text{CM}_{\gamma^* K}
\approx 38$~msr ($\Delta \Omega^\text{CM}_{\gamma^* \pi}
\approx 32$~msr).

Both spectrometers have a QQDQ magnet arrangement with a $45^\circ$
upward bend. The detector packages, placed behind the magnetic
elements, were equipped with four planes of drift chambers for
tracking and two planes of hodoscopes for triggering.  The electron
spectrometer employed a CO$_2$ gas Cherenkov counter and lead glass
shower counters for pion rejection.  For the kaon measurements, clean
particle identification (PID) in the hadron spectrometer was
particularly important because of the very high ratio of $\pi/K$
rates. With the use of two aerogel ($n = 1.015$ and $1.055$) and a
ring-imaging Cherenkov detector (RICH)~\cite{Iodice:2005ex} in the
hadron spectrometer less than 5\% pion contamination in kaon events
was achieved. A negligible pion contamination remained after
additional separation using the coincidence timing between the hadron
and electron signals.  The aerogel counters and coincidence timing
were also effective in removing any proton background.

The measured yields were corrected for detection and reconstruction
efficiencies and dead-time.  Cuts on the events from the PID detectors
were applied to select appropriate particle types, and vertex and
coincidence time cuts were used to reduce background from accidentals.
The missing mass was reconstructed using the measured momenta of the
electron and kaon (pion).

\begin{figure}

\includegraphics[]{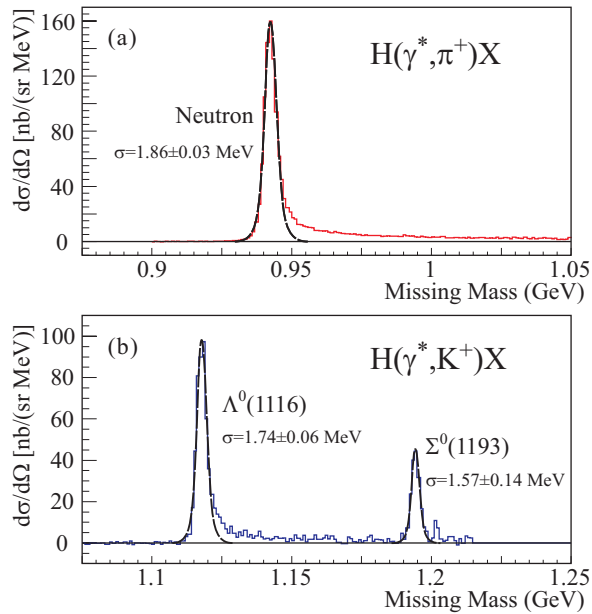}

  \caption{(Color online) The missing mass spectrum for (a)
  $H(\gamma^*,\pi^+)X$ and (b) $H(\gamma^*,K^+)X$.  The dashed curves
  show a fit to the (a) neutron and (b) $\Lambda(1116)$ and
  $\Sigma(1193)$ peaks, from which the missing mass resolution and
  absolute scale uncertainty was determined. \label{fig:mmshift}}

\end{figure}

Calibration data were taken in the $ep\rightarrow e^\prime \pi^+ X$
channel in the missing mass range which included the neutron and in
the $ep\rightarrow e^\prime K^+ X$ channel covering the missing mass
range including the $\Lambda(1116)$ and $\Sigma(1193)$.  The missing
mass resolution was determined by fitting the missing mass peaks for
the neutron, $\Lambda(1116)$ and $\Sigma(1193)$, as shown in
Fig.~\ref{fig:mmshift}.  Based on the mass resolutions of these three
peaks, the momentum resolutions of the left and right spectrometers were
extracted.  From these, the mass resolution in the region of interest
was found to be $\sigma_\text{instr.} = 1.5~\MeVc$.  Also based
on the fit masses of these well-known calibration states, the accuracy
of the reconstructed missing mass was determined to be better than
3~$\MeVc$.

The photoproduction cross section of the $\Lambda(1520)$ was
determined by fitting a Breit-Wigner with an energy dependent
width~\cite{Jackson:1964zd} and a non-interfering background to the
$H(\gamma^*,K^+)X$ missing mass spectra.  From this fit, the cross
section was determined to be $d\sigma / d\Omega\left[\gamma^*
p\rightarrow K^+\Lambda(1520)\right] = 356\pm 25~\text{(stat.)} \pm
35~\text{(syst.)}$~nb/sr with a width of $\Gamma_{\Lambda(1520)} =
16.5 \pm 1.7~\text{(stat.)}~\MeVc$ at $\langle Q^2 \rangle =
0.1~\GeVc^2$.  The largest systematic uncertainty arises from the
absolute acceptance of each of the spectrometer arms for an extended
target. From the fit, the $\Lambda(1520)$ mass was determined to be
$1519.9\pm 0.6~\text{(stat.)}\pm 3~\text{(syst.)}~\MeVc$.  Both the
mass and width are in good agreement with the PDG
averages~\cite{Yao:2006px}.

\begin{figure}

  \includegraphics[]{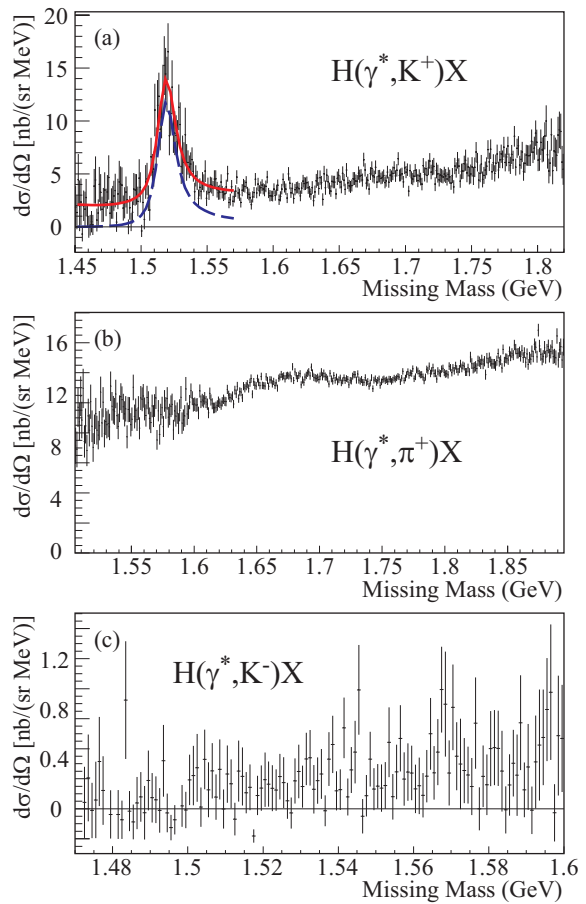}

  \caption{(Color online) The acceptance-weighted, combined missing
  mass spectra obtained for the three reaction channels, after
  accidental coincidence background subtraction: (a)
  $\gamma^*p\rightarrow K^+X$ ($\Sigma^0_5$ search).  The solid red
  curve shows the fit to the $\Lambda(1520)$ and non-resonant
  background and the blue dashed curve shows only the $\Lambda(1520)$
  contribution, (b) $\gamma^*p\rightarrow \pi^+X$ ($\text{N}^0_5$
  search), and (c) $\gamma^*p\rightarrow K^-X$ ($\Theta^{++}$ search).
  Other than the $\Lambda(1520)$ peak in the $K^+$ channel (a) no
  statistically significant, narrow peaks were found in the data.
  \label{fig:missing_mass}}

\end{figure}

For each of the three possible pentaquark states, missing mass spectra
from several (up to eight) different kinematic settings of the
spectrometers were combined.  The individual settings typically
covered a range of approximately 130~MeV in missing mass.  Each
individual missing mass spectrum was transformed into a
photoproduction cross section spectrum in the CM and the accidental
coincidence background was subtracted.  Combining the spectra required
careful relative integrated luminosity normalizations and special
attention to the acceptance weighting as a function of missing mass.
The individual data sets overlapped to some extent, allowing for
verification of the weighting and normalization.  After finalizing the
detector analysis, all transitions between spectrometer settings were
found to be smooth, requiring no {\it ad-hoc} scaling. These spectra
are shown in Fig.~\ref{fig:missing_mass}.

Within each of the search regions, $1550 < M_x < 1810~\MeVc$ for the
$\Sigma^0_5$, $1610 < M_x < 1880~\MeVc$ for the $\text{N}^0_5$ and
$1480 < M_x < 1590~\MeVc$ for the $\Theta^{++}$, the data were
examined for the existence of narrow resonances.  A scan was performed
in 1~$\MeVc$ steps over each search region.  For each mass, $M_x$, and
width, $\Gamma_x$, the goal was to determine a range in the cross
section of a hypothetical resonance that would be compatible with the
missing mass spectrum at 90\% confidence, with no {\it a priori}
assumption of a non-zero resonance cross section.  To accomplish this,
the Feldman and Cousins~\cite{Feldman:1997qc} approach was adopted.
The procedure outlined below was repeated for widths of 0.5, 2, 5 and
10 $\MeVc$.

\begin{figure}

  \includegraphics[]{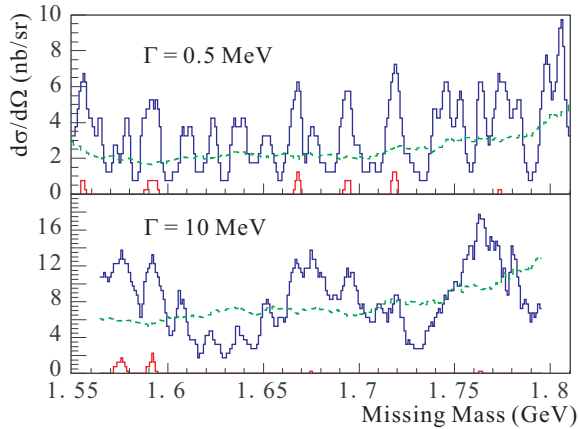}

  \caption{(Color online) The upper and lower limit [top (blue) and
  bottom (red) solid lines] at 90\% confidence level for the
  production of the $\Sigma^0_5$ in the reaction $\gamma^*p
  \rightarrow K^+ \Sigma^0_5$ is shown as function of mass for two
  possible widths of the $\Sigma^0_5$.  In most cases, the lower
  limit (bottom--red curve) is zero.  The smoother green dashed curve
  shows the statistical sensitivity. \label{fig:sigma_cl}}

\end{figure}

\begin{figure}

  \includegraphics[]{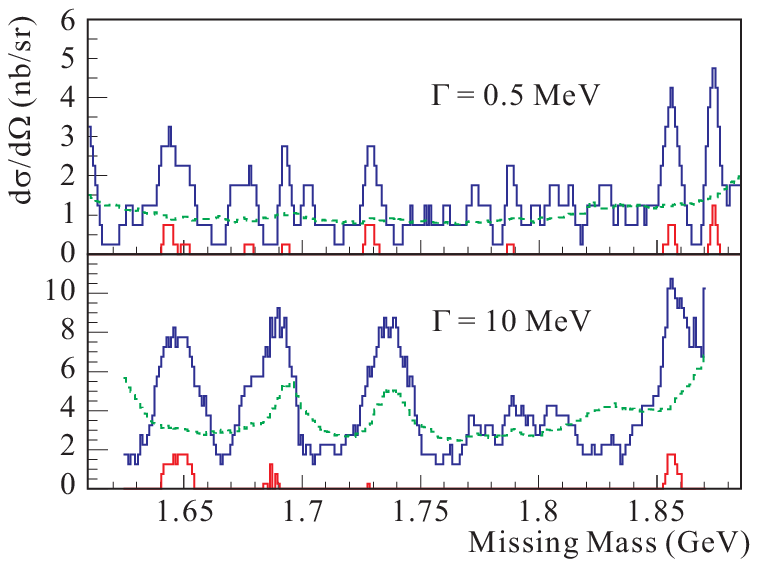}

  \caption{(Color online) The 90\% confidence level limits for the
  production of the $\text{N}^0_5$ in the reaction
  $\gamma^*p\rightarrow \pi^+\text{N}^0_5$ is shown.  The curves have
  the same meaning as in Fig.~\ref{fig:sigma_cl}. \label{fig:n_cl}}

\end{figure}

\begin{figure}[!t]

  \includegraphics[]{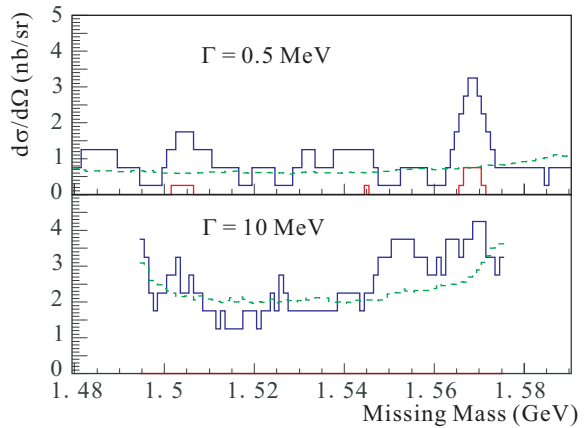}

  \caption{(Color online) The 90\% confidence level for the production
  of the $\Theta^{++}$ in the reaction $\gamma^*p\rightarrow K^-\Theta^{++}$ is
  shown.  The curves have the same meaning as in
  Fig.~\ref{fig:sigma_cl}.  \label{fig:thetapp_cl}}

  \vspace*{-0.3in}

\end{figure}

For a given missing mass, $M_x$, the first step was to establish the
level of background by fitting the 20~$\MeVc$ wide sidebands above and
below the region containing 90\% of a hypothesized peak at $M_x$ with
width $\Gamma_x$.  When $M_x$ was near the edge of the acceptance, the
side band near the edge was reduced, to a minimum of 5~$\MeVc$. With
the background level fixed, the data were fit to a Breit-Wigner at
$M_x$ and width $\Gamma_x$ convoluted with a $1.5~\MeVc$ wide Gaussian
(instrumental resolution) plus the fixed background within the same
window.  The only free parameter in this fit was the cross section of
the Breit-Wigner resonance, $\sigma_\text{best}$, the most
likely cross section, of the possible resonance.  Because of the
paucity of events in some bins for some spectrometer settings, all
fits used maximum likelihood techniques~\cite{Feldman:1997qc,
Baker:1983tu}.  The range in cross section, $\sigma$, accepted at 90\%
confidence was found by examining deviations in the log
likelihood, $\ln{\cal{L}}$, as a function of $\sigma$ from its best
value, $\Delta\ln{\cal{L}}_\text{data}
(\sigma,\sigma^\text{best}_\text{data})$.  The limit in
$\Delta\ln{\cal{L}}$ of the 90\% confidence region was established by
Monte Carlo. For a given \textit{assumed} cross section, $\sigma_x$,
many Monte Carlo ``experiments'' were performed. Missing mass spectra
for each spectrometer setting were randomly populated with total
statistics equal to that of the actual data.  The spectral shape was
based on the smoothed background shape determined in the data analysis
with a hypothetical resonance of cross section $\sigma_x$ added. From
each Monte Carlo ``experiment'', $\Delta
\ln{\cal{L}}_\text{MC}\left(\sigma_x,\sigma^\text{best}_\text{MC}
\right)$, the difference between the hypothetical resonance's cross
section and the best fit of the Monte Carlo spectra was
determined. The distribution of $\Delta\ln{\cal{L}}_\text{MC}$'s from
the Monte Carlo ``experiments'' was examined to determine
$\Delta\ln{\cal{L}}_{90\%}$ such that 90\% of the Monte Carlo
simulations had $\Delta \ln{\cal{L}}_\text{MC}
(\sigma_x,\sigma^\text{best}_\text{MC})
>\Delta\ln{\cal{L}}_{90\%}$\footnote{Recall $\ln{\cal{L}} < 0$ (in the
large statistic limit, $\chi^2 = -2\ln{\cal{L}}$).}.  If $\Delta
\ln{\cal{L}}_\text{data} (\sigma_x,\sigma^\text{best}_\text{data})
>\Delta\ln{\cal{L}}_{90\%}$, then $\sigma_x$ was within the region
accepted with 90\% confidence. In addition, curves indicating the 90\%
probability of background fluctuations were generated using a similar
technique as suggested by Feldman and Cousins~\cite{Feldman:1997qc}.
The upper limit, lower limit and statistical sensitivity curves are
shown in Figs.~\ref{fig:sigma_cl},
\ref{fig:n_cl} and \ref{fig:thetapp_cl} for the $\Gamma_x = 0.5$
and $10~\MeVc$ cases.  The maximum upper limits listed in
Tab.~\ref{tab:uptable} are expressed for each resonance in nb/sr and
as a fraction of the $\Lambda(1520)$ cross section.

\begin{table}

  \caption{This table lists the largest upper limit on the
  photoproduction cross section of the $\Sigma^0_5$, $\text{N}^0_5$
  and $\Theta^{++}$ in nb/sr and relative to the measured
  $\Lambda(1520)$ cross section of $417\pm 30~\text{(stat.)}\pm
  41~\text{syst.}$~nb/sr for resonance widths of $\Gamma$ = 0.5, 2, 5,
  10~$\MeVc$.  \label{tab:uptable}}

  \begin{tabular}{c|*{3}{c@{\extracolsep{5pt}}c@{\extracolsep{20pt}}}}
\hline \hline
$\Gamma$& 
\multicolumn{2}{c}{$\Sigma^0_5$} & 
\multicolumn{2}{c}{$\text{N}^0_5$} & 
\multicolumn{2}{c}{$\Theta^{++}$}\\ 
($\MeVc$) & 
(nb/sr) & \% & 
(nb/sr) & \% &
(nb/sr) & \%  \\ \hline
0.5 & 
10.0 & 2.8 & 
4.5 & 1.3 &
3.0 & 0.8  \\
2.0& 
11.0 & 3.1 & 
5.5 & 1.5 & 
3.5 & 1.0 \\
5.0 & 
13.0 & 3.7 & 
6.0 & 1.7 & 
3.5 & 1.0 \\
10.0 & 
17.5 & 4.9 & 
10.5 & 2.9 & 
4.0 & 1.1 \\ \hline \hline
  \end{tabular}
\end{table}

As can be seen in Figs.~\ref{fig:sigma_cl}, \ref{fig:n_cl} and
\ref{fig:thetapp_cl} most of the 90\% confidence region shows only
upper limits and the upper limit curves oscillate about the
statistical sensitivity curves.  Where the lower limit curves are
different from zero, they are always below the sensitivity curves
implying that none of the lower limits can be distinguished from a
statistical fluctuation.

There are several known or suspected resonances in this mass region,
in particular several 3 or 4-star $\Lambda$ and $\Sigma$ states in the
$\gamma^*p\rightarrow K^+X$ channel~\cite{Yao:2006px}.  Most are too
wide ($>50~\MeVc$) to be visible in this experiment (unless they have
a {\it substantial} cross section) or have only been seen in
partial-wave analyses or both.  Taken together, they add up to a
relatively smooth background.

In conclusion, a high resolution search for the $\Sigma^0_5$,
$\text{N}^0_5$ and $\Theta^{++}$ has been completed using the
Jefferson Lab Hall A HRS spectrometers.  This search had an
instrumental resolution of $\sigma_\text{instr.} = 1.5~\MeVc$.  No
statistically significant narrow ($\Gamma < 10~\MeVc$) structures were
observed in any of the three reaction channels.  Upper limits of the
photoproduction cross section for these states were found to be $<5\%$
of the $\Lambda(1520)$ photoproduction cross section for $\Gamma \le
10~\MeVc$ at 90\% C.L.

We would like to acknowledge the outstanding efforts of the Jefferson
Lab Hall A technical staff that made this work possible.  We are
grateful to W.J. Briscoe, W. Dunwoodie, M. Polyakov and
M. Vanderhaeghen for useful discussions in planning these
measurements.  This work was supported in part by the Italian
Istituto Nazionale di Fisica Nucleare (INFN), Russian State Grant
RSGSS-1124.2003.2, the U.S.-Israeli Binational Science Foundation,
U.S. National Science Foundation and the U.S. Department of Energy,
Office of Nuclear Physics under contracts DE-AC02-06CH11357,
DE-FG02-99ER41110 and DE-AC05-84150, Modification No. M175, under
which the Southeastern Universities Research Association (SURA)
operates the Thomas Jefferson National Accelerator Facility.
\FloatBarrier

\bibliography{pqe}

\end{document}